\documentclass[aip,pop,amsmath,amssymb,preprint]{revtex4-2}
\usepackage{hyperref}
\usepackage{xfrac}
\usepackage{mathtools}
\usepackage[utf8]{inputenc} 
\usepackage{verbatim}
\usepackage{epsfig,graphicx}
\usepackage[english]{babel}

\makeatletter
\newcommand{\fdfrac}[2]{\mbox{\footnotesize$\displaystyle\frac{#1}{#2}$}}
\newcommand\gsl{\ifmmode\textsl{g}\else g\fi}
\newcommand{\sgn}{\operatorname{sign}}
\newcommand{\artanh}{\operatorname{arcth}}
\newcommand{\Si}{\operatorname{Si}}
\makeatother

\begin{document}

\title{Dynamics of equatorial jets generated by temperature fronts}

\author{V.~P.~Goncharov}
\email{v.goncharov@rambler.ru}
\affiliation{A. M. Obukhov Institute of Atmospheric Physics RAS, 109017 Moscow, Russia}


\begin{abstract}
The theory of temperature jets gets extended to account for the influence of the beta effect on their dynamics. Including this effect noticeably changes symmetry properties and laws of conservation inherent to models without the beta effect. Especially nontrivial, the dynamics of jets become near the equator, where the model admits multi-valued solutions that look like jets crossing the equator. For example, the so-called loop solitons moving along the equator with velocities inversely proportional to the cube of their amplitude turn out among them. Estimates based on simple qualitative considerations show that, owing to the beta effect, the temperature jets can form a specific equatorial turbulence, in which they play the role of structural elements. Notably, the spectral slope of the energy density at large wave numbers in such turbulence becomes equal to unity.
\end{abstract}

\maketitle

\section{Introduction}\label{sec1}
Frontal jet currents in the ocean and the atmosphere are caused by horizontal pressure gradients that in rotating nonhomogeneous fluids occur due to having narrow transition zones between regions of different potential vorticity (PV) and/or temperature under the action of the Coriolis and gravity forces~\cite{cb11}. With varying degrees of detail, this effect is reproduced by many models of geophysical hydrodynamics. But, most simply, it can be mathematically formalized and studied in the context of vertically averaged quasi-geostrophic models.

One such approach can be developed within the thermal rotating shallow-water (TRSW)~\cite{rip93,rip95,wd13} model, which describes a two-dimensional (columnar) motion of incompressible horizontally-nonuniform fluid in a sufficiently thin layer under the shallow-water approximation and in neglecting any dissipation. In the TRSW model, two scenarios of jet generation are possible, depending on which field (potential vorticity or temperature) undergoes the jump at the frontal interface.

The theory of thin jets initiated by cross-frontal PV jumps can be developed only assuming the horizontal homogeneity of the temperature field. Otherwise, the potential vorticity has no Lagrangianity property, and it is this property that allows us to trace the jet's trajectory during its evolution. A brief description of some models and theoretical approaches~\cite{rn67,lf75,fr84,ps86,p88,nds93,rp94,cpm97} that have been used at different times to study the trajectory dynamics of jets generated at PV fronts can be found in Ref.~\cite{cpr93}.

An alternative scenario of thin jets resulting from temperature jumps has been recently analyzed in Ref.~\cite{g21}. Like PV jets, temperature jets also have a Lagrangian character, but their trajectories are localized not at lines of PV jumps but at temperature jump lines. Although, in deriving, both theories assume the same relation $L~T^{1/3}$ between the characteristic scales of length $L$ and time $T$, their equations of motion, symmetries, and hence solutions turn out to be fundamentally different.

Another essential difference between the theories exhibits when accounting for the beta effect. If, for the PV jets, the inclusion of the beta effect leads to a Doppler shift without any violation of the translational invariance of the theory, for the temperature jets, it is not so at all. Moreover, in this case, their path equations lose, as well, some other symmetries to be inherent for the planar version of the theory when $\beta=0$. For instance, path equations become no longer invariant under rotations and inversions. Instead of them, the theory gains the properties that open possibilities for existing solutions of a new type. These solutions, known as loop solitons~\cite{ps86,rp94,cpr93}, destroy the connectedness of the flow by breaking it into closed domains and can enhance large-scale zonal transport.

One more application area for the concept of temperature jets is large-scale turbulence. That means the collective contribution of temperature jets to atmospheric dynamics can be considered a statistical process in which they play the role of structural elements. In this respect, studying the spectral properties of equatorial turbulence and investigating the beta-effect influence on temperature jets are two sides of the same coin.

The paper is structured as follows. In Sec.~\ref{sec2}, we outline a general statement of the problem and formulate the variational principle for a depth-averaged, horizontally inhomogeneous layer of a rotating fluid. In Sec.~\ref{sec3}, by using the scaling of variables, we develop a self-consistent perturbation theory to find the thin jets Lagrangian in the equatorial beta-plane approximation. Sec.~\ref{sec4} analyzes the motion equations describing the dynamics of equatorial jets and includes some solutions, among which loop solitons. In Sec.~\ref{sec5}, we discuss obtained results. The validity of the equatorial parameterization for Coriolis is argued in Appendix~\ref{sec6} in the context of its applicability to a layer on a rotating sphere. Appendix~\ref{sec7} addresses the spectral properties of turbulence for which straight-line temperature jets are its structural elements.

\section{General statement}\label{sec2}

The shortest path to formulate the equations of motion for frontal temperature jets, with the account of the beta effect, is to use the variational least action principle
\begin{equation}
\delta S=0,\quad S=\int\mathcal{L}dt,\label{eq:1}
\end{equation}
where $S$ is the action and $\mathcal{L}$ is the Lagrangian, which, based on physically reasonable approximations, needs to be successively simplified up to its final form.

As a first step towards this goal, we consider the model of a two-layer inviscid incompressible fluid that moves under the action of gravity and Coriolis forces and occupies the half-space $z>0$, where $z$ is the vertical coordinate. Let the lower layer of a sufficiently small height $h$ rest on the impermeable bottom $z=0$ and be horizontally inhomogeneous with the density $\varrho_{0}+\Delta\varrho$. But in doing so, the upper layer, filled with the homogeneous fluid of the density $\varrho_{0}$, spans to infinity. Moreover, if all kinetic energy is assumed to be determined mainly by horizontal motions and concentrated in the lower layer above which there is a weakly perturbed, almost homogeneous fluid, then, after integration over $z$, we obtain the Lagrangian
\begin{equation}
\mathcal{L}=\frac{1}{2}\int h\left(u^{2}+v^{2}-
2uR-hb\right)dxdy.\label{eq:2}
\end{equation}

In this expression, besides those already specified, we use the following notations: $x$, $y$ are the horizontal Cartesian coordinates, $t$ is the time, subscripts denote the partial derivatives, $u$, $v$ are components of the horizontal velocity. Responsible for rotation, the function $R$ depends only on $y$, and its derivative gives the Coriolis parameter $f$:
\begin{equation}
R_{y}=f.\label{eq:3}
\end{equation}

To account for the influence of Earth's sphericity, we will use the so-called equatorial beta-plane approximation
\begin{equation}
f=\frac{2\Omega}{a}y,\label{eq:4}
\end{equation}
where $\Omega$ is the angular rotation rate, and $a$ is the planetary radius. Integrating this parameterization over $y$ leads to the expression
\begin{equation}
R=\frac{\Omega}{a}y^{2}.\label{eq:5}
\end{equation}

As well known, the beta effect is one of the keystones in geophysical fluid dynamics. For the model under study, that is quite explainable since the presence of the equator removes some symmetries from the dynamics of the jets, which are inherent to them in the planar case. Notably, if we consider the above model on a rotating sphere and use Lambert's cylindrical projection (see Appendix~\ref{sec6} for more details)
\begin{equation}
y=a\sin\varphi\label{eq:6}
\end{equation}
to introduce the variable $y$ instead of the latitude $\varphi$, then, after such the transformation, the parameterization (\ref{eq:5}) becomes exact.

At last, the latter term in Lagrangian (\ref{eq:2}) is responsible for potential energy and includes the variable $b$ denoting the relative buoyancy. When the deviations of density and temperature are related linearly, it is possible to determine this quantity alternatively as
\begin{equation}
b=\gsl\Delta\varrho/\varrho_{0}=
-\gsl\Delta\theta/\theta_{0},\label{eq:7}
\end{equation}
where $\gsl$ is gravity, $\Delta\varrho(x,y,t)$ and $\Delta\theta(x,y,t)$ are the density and temperature deviations from the respective reference values $\varrho_{0}$ and $\theta_{0}$.

Depending on which is the description of the fluid continuum, there are mainly two ways to apply the principle of least action to similar types of models. One of these ways relies on the Eulerian conception of fluid motion and uses Lin's constraints, while the other, called Lagrangian, requires tracking individual fluid particles.

In the context of depth-averaged fluid models, a pivotal point of such an approach is to interpret flow as the motion of infinitely narrow liquid columns whose Cartesian positions
\begin{equation}
x=\hat{x}(s,n,t),\quad y=\hat{x}(s,n,t)\label{eq:8}
\end{equation}
are functions of time $t$ and labeling coordinates $(s,n)$ of the special kind. Being assigned so that equal infinitesimal areas in $(s,n)$-space correspond to the same volumes, these coordinates, sometimes termed straightened~\cite{zeit18}, obey the condition
\begin{equation}
Hdsdn=hdxdy,\label{eq:9}
\end{equation}
in which $H$ is a constant and means the average layer depth.

Since, apart from (\ref{eq:9}), the transformation (\ref{eq:8}) entails the rules both for velocities
\begin{equation}
u\rightarrow\hat{x}_{t},\quad v\rightarrow\hat{y}_{t},\label{eq:10}
\end{equation}
and also for buoyancy and layer height
\begin{equation}
\quad b\rightarrow B(s,n),\quad h\rightarrow H/J(\hat{x},\hat{y}),\label{eq:11}
\end{equation}
where $J$ is the transform Jacobian
\begin{equation}
J(\hat{x},\hat{y})=\hat{x}_{s}\hat{y}_{n}-
\hat{x}_{n}\hat{y}_{s},\label{eq:12}
\end{equation}
we find that the Lagrangian (\ref{eq:2}) written in terms of straightened variables looks as
\begin{equation}
\mathcal{L}=\frac{H}{2}\int\left(\hat{x}_{t}^{2}+
\hat{y}_{t}^{2}-2\frac{\Omega}{a}\hat{x}_{t}\hat{y}^{2}-
\frac{HB}{J(\hat{x},\hat{y})}\right)dsdn.\label{eq:13}
\end{equation}

\section{Scaling and perturbation theory on equatorial beta-plain}\label{sec3}

In any geophysical context, the theory of thin jets has an approximate character and, as a rule, for one's construction, requires the use of the multiscale expansions method. Because of this, to more effectively apply perturbation theory and to control all small parameters, the best way is to convert the Lagrangian (\ref{eq:13}) to a dimensionless form.

For frontal jets caused by temperature gradients and formed by the influence of gravity and rotation, it is natural to take the following quantities
\begin{equation}
T=\frac{1}{2\Omega},\quad L=T\left(\gsl H\frac{\Delta\theta^{*}}{\theta_{0}}\right)^{1/2},\label{eq:14}
\end{equation}
as the typical scales of time and horizontal length, respectively. Here $H$ denotes the layer depth reached far from the jet in a quiescent region, and $\Delta\theta^{*}$ is the peak value of temperature deviation. In particular, in conditions typical of the Earth's atmosphere, where
\begin{equation}
H\approx5\;km,\quad\gsl\approx10\;m\cdot s^{-2},
\quad\Omega\approx7.3\cdot10^{-5}\;s^{-1},
\quad \Delta\theta^{*}/\theta_{0}\approx5\cdot10^{-3},\label{eq:15}
\end{equation}
the definition (\ref{eq:14}) allows us to estimate the Rossby length scale as $L\approx100\;km$.

Now, upon transforming by the rules
\begin{gather}
\left(\hat{x},\hat{y},s,n\right)\rightarrow
L\left(\hat{x},\hat{y},s,n\right),\quad t\rightarrow Tt,\label{eq:16}\\
B\rightarrow B\frac{L^{2}}{T^{2}H},\quad
\mathcal{L}\rightarrow\mathcal{L}\frac{L^{4}}{T^{2}},\label{eq:17}
\end{gather}
we can reduce the Lagrangian (\ref{eq:13}) to the dimensionless form
\begin{equation}
\mathcal{L}=\frac{1}{2}\int\left(\hat{x}_{t}^{2}+\hat{y}_{t}^{2}-
\delta\hat{x}_{t}\hat{y}^{2}-\frac{B}{J(\hat{x},\hat{y})}\right)dsdn.\label{eq:18}
\end{equation}
Here $\delta=L/a<1$ is the parameter responsible for the beta effect.

Let the jet be sufficiently thin and localized at the zero-buoyancy line described parametrically as
\begin{equation}
x=X(s,t),\quad y=Y(s,t).\label{eq:19}
\end{equation}

The thin-jets approximation implies an expansion of the field variables in the vicinity of the jet trajectory, and it assumes that the dependent and independent variables appearing in the Lagrangian (\ref{eq:18}) have the following orders:
\begin{equation}
s\sim\hat{x}\sim\hat{y}\sim1/\epsilon,\quad
n\sim B\sim1,\quad t\sim1/\lambda,\label{eq:20}
\end{equation}
where $\epsilon$ and $\lambda$ are two small perturbation parameters. How they relate to each other and the parameter $\delta$ will be found later. Note that, although the smallness of $\epsilon$ is arbitrary, it will be measured as the ratio $\epsilon\sim l/L$, where $l$ is the width of the jet.

Then, expanding the functions $\hat{x}$ and $\hat{y}$ by powers of $n$ and $\epsilon$, one can find that
\begin{gather}
\hat{x}=\frac{1}{\epsilon}\left(X+\epsilon nX_{1}+\epsilon^{2}n^{2}X_{2}+O(\epsilon^{3})\right),\label{eq:21}\\
\hat{y}=\frac{1}{\epsilon}\left(Y+\epsilon nY_{1}+\epsilon^{2}n^{2}Y_{2}+O(\epsilon^{3})\right),\label{eq:22}
\end{gather}
where $X_{1}(s,t)$, $Y_{1}(s,t)$ are the displacements in the first order of perturbation theory, and $X_{2}(s,t)$, $Y_{2}(s,t)$ are the displacements in the second order.

The expansion for buoyancy we can find analogously. The fact that along-jet velocities localize near zero-buoyancy lines and reach their peak values therein allows us to conclude that the function $B(s,n)$ must be odd relative to $n$, so its expansion must begin with a linear term
\begin{equation}
B=n\gamma+O(\epsilon^{2}).\label{eq:23}
\end{equation}
Here the coefficient $\gamma(s)$ characterizing the across-jet temperature gradient (or the along-jet velocity) can depend only on the longitudinal coordinate $s$.

The Jacobian $J$ is one more quantity that we should expand into a power series in $\epsilon$. Substituting expansions (\ref{eq:21}) and (\ref{eq:22}) into the Jacobian (\ref{eq:12}), it is easy to find
\begin{gather}
J=J_{0}+n\epsilon J_{1}+O(\epsilon^{2}),\label{eq:24}\\
J_{0}=X_{s}Y_{1}-Y_{s}X_{1},\label{eq:25}\\
J_{1}=X_{1s}Y_{1}-Y_{1s}X_{1}+2\left(X_{s}Y_{2}-Y_{s}X_{2}\right).\label{eq:26}
\end{gather}

Now, in analogy with Ref.~\cite{g21}, we can find the thin-jet Lagrangian in the equatorial beta-plane approximation. By inserting the expansions (\ref{eq:21})--(\ref{eq:26}) directly into (\ref{eq:18}), integrating it over $-1\leq n\leq 1$, and keeping the expansion terms up to the first order of $\epsilon$, $\lambda$, and $\delta$ we arrive at the expression
\begin{equation}
\mathcal{L}=\int\Biggl(-\frac{\lambda\delta}{\epsilon^{3}}X_{t}Y^{2}+
\frac{\epsilon}{3}\gamma\frac{J_{1}}{J_{0}^{2}}\Biggr)ds+O(\epsilon,\lambda).\label{eq:27}
\end{equation}

To obtain a closed description in the leading order of perturbation theory, we need to require that both terms in the integral (\ref{eq:27}) have the same order. Whence, by comparing them, we get
\begin{equation}
\lambda=\frac{\epsilon^{4}}{3\delta}.\label{eq:28}
\end{equation}
In addition, following Ref.~\cite{g21}, we have to impose the conditions
\begin{gather}
X_{s}X_{1}+Y_{s}Y_{1}=0,\quad X_{s}Y_{2}-Y_{s}X_{2}=0,\label{eq:29}
\end{gather}
which say that first-order displacements are orthogonal to the jet trajectory while second-order ones are parallel. Note that these conditions provide self-consistency of the perturbation theory so that it would not contain any second-order displacements in the first-order perturbations for the Jacobian $J$.

With the use of the relation (\ref{eq:25}) and the orthogonality condition from (\ref{eq:29}), one can find
\begin{equation}
X_{1}=-\frac{J_{0}Y_{s}}{X_{s}^{2}+Y_{s}^{2}},\quad
Y_{1}=\frac{J_{0}X_{s}}{X_{s}^{2}+Y_{s}^{2}},\quad
J_{1}=J_{0}^{2}\frac{Y_{s}X_{ss}-X_{s}Y_{ss}}
{\left(X_{s}^{2}+Y_{s}^{2}\right)^{2}}.\label{eq:30}
\end{equation}
As a result, after excluding $J_{1}$ from the Lagrangian (\ref{eq:27}), it reduces to a more simple form
\begin{equation}
\mathcal{L}=\int\Biggl(-X_{t}Y^{2}+\gamma\frac{Y_{s}X_{ss}-
X_{s}Y_{ss}}{\left(X_{s}^{2}+Y_{s}^{2}\right)^{2}}\Biggr)ds.\label{eq:31}
\end{equation}

\section{Path equations, some symmetries and solutions}\label{sec4}

Formulated based on the variational principle (\ref{eq:1}) with the Lagrangian (\ref{eq:31}), the equations of equatorial dynamics for temperature jets already externally look different than those derived for a planar version of this theory~\cite{g21}. Formally, the equatorial version follows directly from the planar one, at once after substituting $X_{t}\rightarrow X_{t}Y$ and $Y_{t}\rightarrow Y_{t}Y$ into the left-hand sides of planar version equations. In particular, when the temperature gradient $\gamma$ across the jet is assumed to be constant, this trick leads us to the path equations
\begin{equation}
X_{t}Y=\frac{\partial}{\partial s}\left(\frac{X_{ss}}
{\left(X_{s}^{2}+Y_{s}^{2}\right)^{2}}\right),\quad
Y_{t}Y=\frac{\partial}{\partial s}\left(\frac{Y_{ss}}
{\left(X_{s}^{2}+Y_{s}^{2}\right)^{2}}\right),\label{eq:32}
\end{equation}
whose symmetries are essentially different from those inherent to the planar version.

One can easily check that Eqs.~(\ref{eq:32}) lose some of those symmetries, due to which they remained invariant in the planar version. For example, they become no invariant under translations in $Y$, rotations, and inversions. Instead, Eqs.~(\ref{eq:32}) gain new symmetry properties
\begin{gather}
X\rightarrow X+X_{0},\label{eq:33}\\
X(s,t)\rightarrow X(-s,t),\quad Y(s,t)\rightarrow -Y(-s,t).\label{eq:34}
\end{gather}

From all the integrals of motion found for the planar version~\cite{g21}, the equatorial version (\ref{eq:32}) leaves invariant only the energy integral
\begin{equation}
E=\frac{1}{2}\int\frac{Y_{ss}X_{s}-X_{ss}Y_{s}}
{\left(X_{s}^{2}+Y_{s}^{2}\right)^{2}}ds.\label{eq:35}
\end{equation}
But additionally, it provides time invariance of two more integrals:
\begin{equation}
I_{1}=\int Y^{2}ds,\quad I_{2}=\int X_{s}Y^{2}ds.\label{eq:36}
\end{equation}

There is a fundamental difference between how the beta effect impacts the dynamics of temperature jets and how it influences the motion of PV jets generated by potential vorticity jumps. While, for the PV jets~\cite{cpr93}, the beta-effect leads to Doppler's shift of the phase velocity, its influence on temperature jets turns out to be more nontrivial.

Unfortunately, it is currently unknown whether the Eqs.~(\ref{eq:32}) are exactly integrable. But despite this, they admit a wide variety of steadily propagating solutions, including spatially periodic waves. Herewith, the beta-effect influence manifests itself most brightly in the soliton-type solutions. In particular, due to symmetry (\ref{eq:34}), the soliton solutions for equatorial jets are not unipolar. The sign of their amplitude depends on their motion direction. Another remarkable feature of temperature jets is, owing to the beta effect, they can evolve in the shape of loop solitons (both with- and non-intersection). As known in geophysical fluid dynamics~\cite{ps86,rp94}, the solutions of such type can be responsible for the fluid detachment from jet flows and are used to explain the formation of warm/cold rings in the Gulf Stream.

Consider the solutions of Eqs.~(\ref{eq:32}) within a specific class. Quite obviously, in the case $X=s$, i.e., when $s$ coincides with the axis $X$, these equations reduce to one single
\begin{equation}
Y_{t}Y=\frac{\partial}{\partial X}\frac{Y_{XX}}
{\left(1+Y_{X}^{2}\right)^{2}},\label{eq:37}
\end{equation}
with the variable $Y$ to be considered as a function of $X$ and $t$.

Note that being scale-invariant under the transformation
\begin{equation}
\left(X,Y,t\right)\rightarrow
\left(\alpha X,\alpha Y,\alpha^{4}t\right),\label{eq:38}
\end{equation}
with rescaling factor $\alpha$, Eq.~(\ref{eq:37}), in particular, conserves the integrals
\begin{equation}
Q_{1}=\int Y^{2}dX,\quad Q_{2}=\int XY^{2}dX,\quad
Q_{3}=\int Y^{3}dX.\label{eq:39}
\end{equation}

Let us look for the steadily translating solutions $Y(\sigma)$, $(\sigma=X-ct)$, propagating with a constant velocity $c$ without profile deformation along the $X$ axis. Then, after twofold integration, scaling, and the introduction of the calibration parameter $\kappa$
\begin{equation}
y=\frac{Y}{Y_{0}},\quad x=\frac{\sigma}{Y_{0}},\quad \kappa=\frac{3}{cY_{0}^{3}},\label{eq:40}
\end{equation}
which links the amplitude $Y_{0}$ and phase velocity $c$, Eq.~(\ref{eq:37}) can be reduced to the form
\begin{equation}
y_{x}^{2}=\frac{-P}{\kappa+P},\quad P=y^{3}+c_{1}y+c_{2}.\label{eq:41}
\end{equation}
Here and below, $P$ is an incomplete cubic polynomial, and $c_{1}$ and $c_{2}$ are integration constants chosen to satisfy the appropriate boundary conditions at infinity.

\subsection{Cusped jets}

Limiting ourselves to the so-called solitary solutions, we first consider those to have a power decay at infinity. One can show that such solutions become possible if $c_{1}=c_{2}=0$. In this case, assuming without any loss of generality $\kappa=-1$, Eq.~(\ref{eq:41}) takes the form
\begin{equation}
y_{x}^{2}=\frac{y^{3}}{1-y^{3}}.\label{eq:42}
\end{equation}

Solutions of this equation vary in the range $0<y\leq1$ and, being translational invariant, depend on $x$ with the accuracy of an additive constant omitted next for brevity. In implicit form, the general solution of (\ref{eq:42}) can be written as
\begin{equation}
x=\pm B(\fdfrac{5}{6},\fdfrac{1}{2})\left(1-y^{-1/2}
P_{\frac{1}{2}}^{\left(-\frac{1}{6},-\frac{3}{2}\right)}
\left(1-2y^3\right)\right),\label{eq:43}
\end{equation}
through the fractional Jacobi function~\cite{gbl05}. To find the phase velocity $c$ of such solutions, one has to use the third relation from (\ref{eq:40}), whence, at $\kappa=-1$, it follows the dispersion relation
\begin{equation}
c=-3/Y_{0}^{3},\label{eq:44}
\end{equation}
for their phase velocity.

\begin{figure}[t!]
\centerline{\includegraphics[width=\columnwidth]{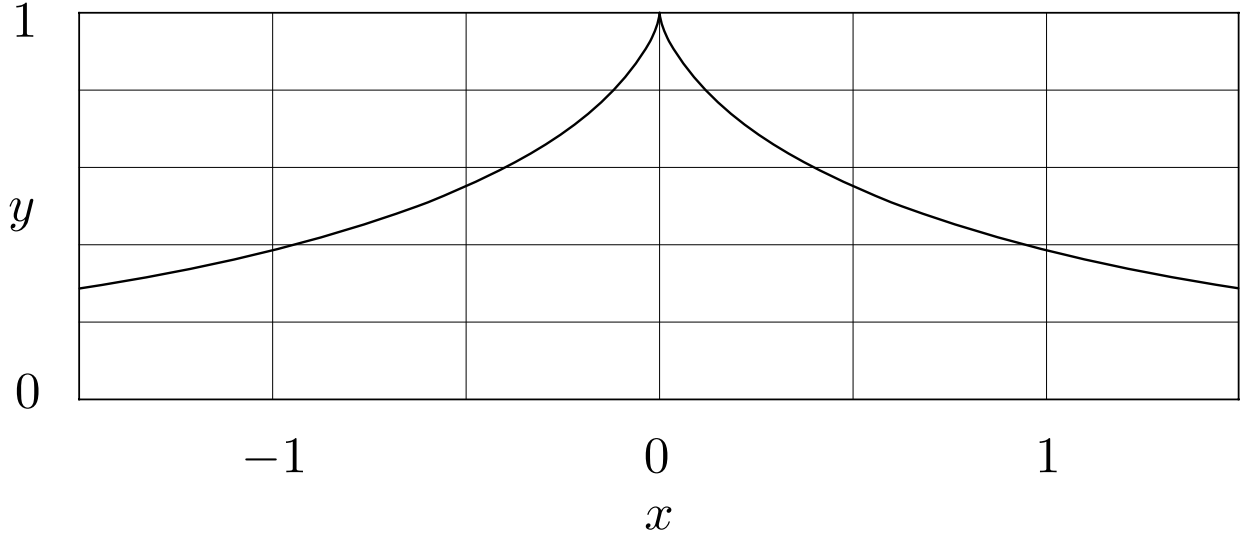}}
\caption{The cusped-jet solution.}\label{fig1}
\end{figure}
Note that found solutions are cusped (see Fig.~\ref{fig1}). At infinity, they vanishes by the power law $y\propto x^{-2}$, but at peak point, $x=0$, they take the finite value $y=1$ and near it behave like $1-y\propto x^{2/3}$. Such behavior of cusped solutions cause their derivatives at these points to have power asymptotics
\begin{equation}
y_{x}\bigr|_{x\rightarrow 0}\propto x^{-1/3},\quad
y_{x}\bigr|_{x\rightarrow\infty}\propto x^{-3}.\label{eq:45}
\end{equation}

\subsection{Loop jets}

Another class of steady-state solutions admissible for Eq.~(\ref{eq:41}) are the so-called solitons. Given their specific behavior, we will assume that they reach a maximum of $y=1$ at $x=0$, while, at infinity, exponentially decrease as $e^{-|x|/\lambda}$, approaching there a constant value. A simple analysis shows that such solutions are possible only if $c_{1}=-3/4$, $c_{2}=-1/4$, and imply the fulfillment of the following conditions:
\begin{equation}
P=(y-1)\left(y+\frac{1}{2}\right)^{2},\quad
-\frac{1}{2}\leq y\leq 1,\quad \kappa=\frac{3}{2}\lambda^{2}\geq\frac{1}{2}.\label{eq:46}
\end{equation}
Here the last equality can be viewed as a relation between the parameter $\kappa$ and the length scale $\lambda$ that defines how strongly localized the solution is. If we take advantage of the proper equalities in Eqs.~(\ref{eq:40}) and (\ref{eq:46}), then excluding $\kappa$ from them leads us to the dispersion relation
\begin{equation}
c=2\lambda^{-2}Y_{0}^{-3}.\label{eq:47}
\end{equation}
It demonstrates that the phase velocity for solitons depends not only on their amplitude, as in the case of cuspons, but also on their localization length.

Thus, with the account of (\ref{eq:46}) and after replacing
\begin{equation}
y=\cos\psi,\quad x=x(\psi),\quad |\psi|\leq2\pi/3,\label{eq:48}
\end{equation}
Eq.~(\ref{eq:41}) looks like
\begin{equation}
x_{\psi}^{2}=\sin^{2}\psi\left[2\kappa\csc^{2}
\left(\frac{3\psi}{2}\right)-1\right].\label{eq:49}
\end{equation}

\begin{figure}[t!]
\centerline{\includegraphics[width=\columnwidth]{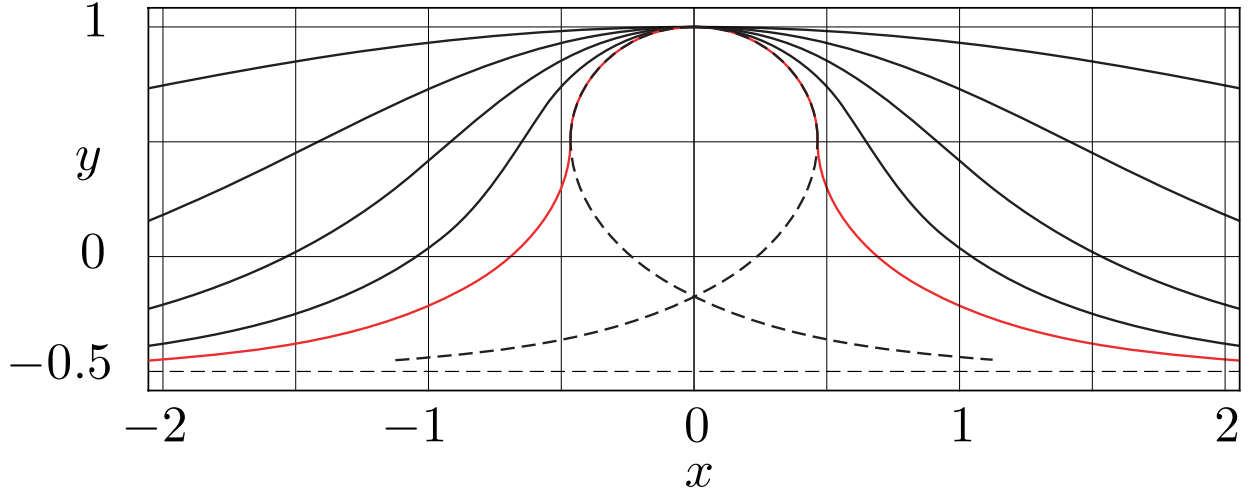}}
\caption{Single-valued solitons for the sequentially increasing parameter $\kappa$, running values $0{.}5$, $0{.}65$, $1$, $2$, $8$. As $\kappa$ increases, the solitons decrease with distance more and more slowly. The limiting value $\kappa=0{.}5$ corresponds to the red curve. The dashed curve corresponds to the loop soliton with the intersection.}\label{fig2}
\end{figure}

As shown in Fig.~\ref{fig2}, this equation has a family of one-valued solitary solutions only under the condition $\kappa>1/2$.
The multi-valued solitary solutions arise in the limiting case $\kappa=1/2$ when Eq.~(\ref{eq:49}) takes the form
\begin{equation}
x_{\psi}^{2}=
\sin^{2}\psi\cot^{2}\left(\frac{3\psi}{2}\right).\label{eq:50}
\end{equation}
In this instance, as we shall see below, it is integrated analytically. However, the final result will depend on choosing the so-called signum function that arises in extracting the square root.

To begin with, let us consider the simplest multi-valued soliton solution constructed without using any signum functions:
\begin{gather}
x=F(\psi),\quad y=\cos\psi,\label{eq:51}\\
F(\psi)=\frac{2}{\sqrt{3}}\artanh\left(\frac{1}{\sqrt{3}}
\tan\left(\frac{\psi}{2}\right)\right)-
\sin\psi.\label{eq:52}
\end{gather}
Here $F(\psi)$ is the particular solution of Eq.~(\ref{eq:50}) which, as shown in Fig.~\ref{fig2}, corresponds to the so-called loop soliton with self-intersection.

If we want to find loop solitons free of self-intersections, we must construct the relevant signum functions
\begin{equation}
S_{+}=\sgn\left(\psi+\frac{\pi}{3}\right),\quad
S_{-}=\sgn\left(\psi-\frac{\pi}{3}\right),\label{eq:53}
\end{equation}
whose zeros $\psi=\pm\pi/2$ coincide with the zeros of the derivative $dF/d\psi$. As a result, we get three possible solutions
\begin{equation}
x=x_{\pm}=S_{\pm}\left(F(\frac{\pi}{3})\pm F(\psi)\right),\quad
x=F(\psi)-x_{+}-x_{-}.\label{eq:54}
\end{equation}

\begin{figure}[h]
\centerline{\includegraphics[height=12cm]
{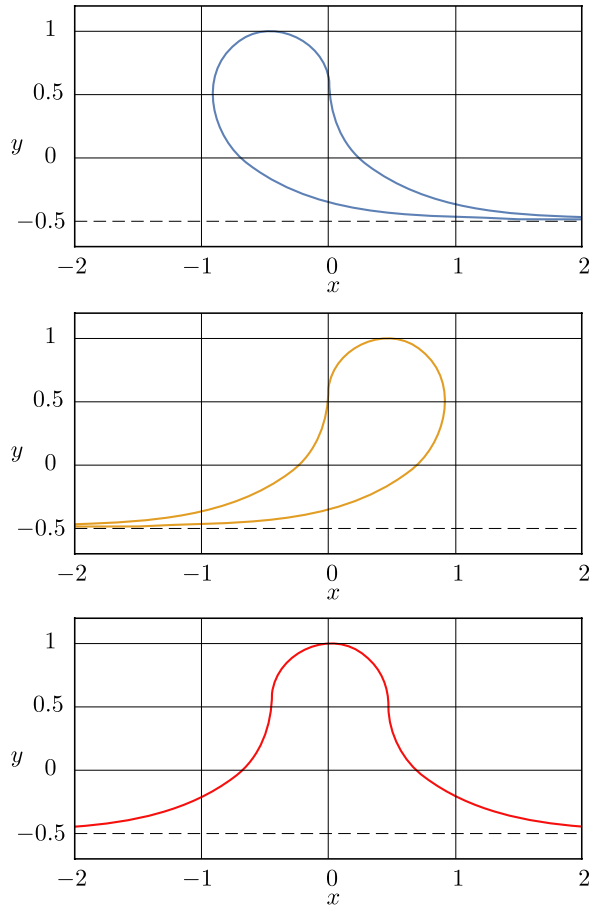}}
\caption{The loop solitons without self-intersection. The blue curve corresponds to the $x_{+}$-solution, the yellow curve to the $x_{-}$-solution, and the red curve to the $x$-solution of Eq.~(\ref{eq:50}).}\label{fig3}
\end{figure}

Since differentiating signum functions over $\psi$ leads to delta-functions, it is easy to check that all solutions listed in (\ref{eq:54}) and illustrated in Fig.~\ref{fig3} satisfy Eq.~(\ref{eq:50}). Note also that, unlike the first two solutions, the last solution is one-valued and coincides with (\ref{eq:52}). Figs.~\ref{fig2} and~\ref{fig3} show it in red.

\section{Discussion and conclusions}\label{sec5}

In this paper, based on the variational principle formulated for the theory of temperature-frontal jets, the planar version~\cite{g21} of this theory gets extended to include the beta effect. In contrast to jets on potential vorticity jumps~\cite{cpr93}, where the beta effect results only in a Doppler shift, for jets on temperature jumps, we found that the beta effect leads to the fundamental changes of dynamical symmetries in the model. As a result, the corresponding path equations admit a new class of solutions. In particular, as our analysis has shown, among them turn out equatorial jets in the form of cuspons and loop solitons.

As follows from (\ref{eq:44}) and (\ref{eq:47}), the phase velocities of both solutions have the same dependence on the amplitude and, being opposite in sign, are in the relation $2|c_{sol}|\leq|c_{cusp}|$, where equality is achieved on loop solitons. Based on conditions (\ref{eq:15}) the velocity estimations at $Y_{0}\approx L\approx 100\;km$ give us the values $|c_{cusp}|\approx 6\;m\cdot s^{-1}$ and $|c_{loop}|\approx 12\;m\cdot s^{-1}$, respectively for cusped and loop solutions.

Unfortunately, due to the intense mixing and turbulence in the real atmosphere, these solutions are almost impossible to identify with sufficient reliability in experimental in-situ data or direct satellite observations. However, if we assume that some of these solutions can play the role of structural elements of large-scale turbulence, this hypothesis can be tested by indirect methods, both analytically and numerically.

One way to implement this approach is to consider large-scale turbulence as a collective effect provided by a statistical ensemble constructed from regular solutions found analytically. In so doing, the free parameters of such solutions should be treated as random variables. It is worth noting that similar approaches have been applied in Refs.~\cite{ph58,saf71,kp73,kuz04,knnr07,gp14} but in other contexts.

\begin{figure}[t]
\centerline{\includegraphics[width=0.6\columnwidth]{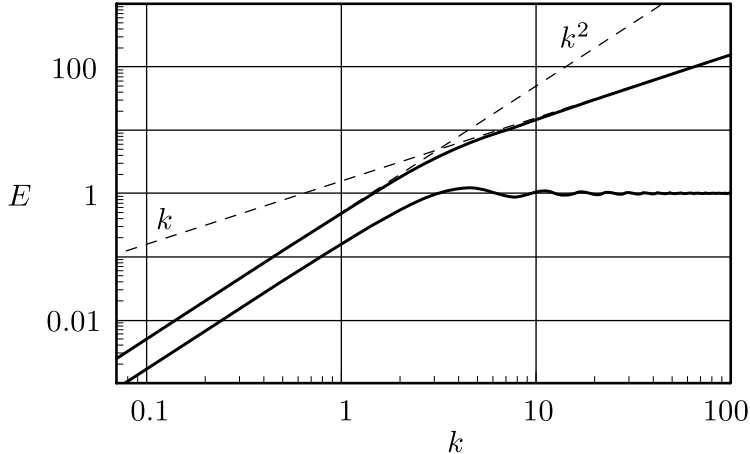}}
\caption{The energy spectrum of turbulence created by equatorial temperature jets. The dashed lines on the log-log plot show power asymptotes. In the lower plot, the equatorial beta effect is off.}
\label{fig4}
\end{figure}
In the spirit of these ideas, it can be demonstrated that, at the expense of equatorial temperature jets, the beta effect may be responsible for modifying the spectral properties of turbulence in the near equator. In particular, as shown in Appendix~\ref{sec7} (see also Fig.~\ref{fig4}), even neglecting trajectory dynamics of temperature jets, their impact on the spectral slope in the short-wave range can be sufficiently strong.

\acknowledgments
This work was supported by the Russian Science Foundation, 23-17-00273.

\appendix
\section{Temperature jets on rotational sphere}\label{sec6}

Because inertial effects play a secondary role in the dynamics of thin temperature jets, to extend the theory to a rotational sphere, let us start at once with the truncated Lagrangian
\begin{equation}
\mathcal{L}=\frac{a^{2}}{2}\int h\left(2a^{2}\Omega \frac{d\vartheta}{dt}\cos^{2}\varphi-
hb\right)\cos\varphi d\varphi d\vartheta,\label{eq:A1}
\end{equation}
ignoring all other effects except Coriolis and buoyancy. Here, $a$ is the radius of the sphere, $h$ is the thickness of an atmospheric layer on it, $\varphi$ and $\vartheta$ are the geographic coordinates (latitude $|\varphi|\leq\pi/2$ and longitude $|\vartheta|\leq\pi$), $\Omega$ is the rotational angular speed, $b$ is the relative buoyancy. In the Eulerian description, the dynamic variables $h$, $b$, and the azimuthal velocity $d\vartheta/dt$ are treated as functions of the spherical coordinates, $\varphi$ and $\vartheta$, and time $t$.

The Lagrangian (\ref{eq:A1}) describes depth-averaged (columnar) motions of an incompressible fluid with variable buoyancy that covers the rotating sphere by a thin layer ($h\ll a$) and is under the action of gravity and Coriolis forces. To formulate the variational principle similar to (\ref{eq:1}), we must free ourselves from constraints imposed implicitly in the Eulerian formulation. To do this, let us pass from the Eulerian description to the Lagrangian one, considering the geographic coordinates of fluid columns
\begin{equation}
\vartheta=\hat{\vartheta}(s,n,t),\quad\varphi=\hat{\varphi}(s,n,t),\label{eq:A2}
\end{equation}
as functions of curvilinear labeling coordinates, $s$ and $n$, and time $t$.

Since the motion of such a fluid is assumed incompressible and without dissipation, each infinitesimal fluid column must preserve its volume and buoyancy. So, the Lagrangian (\ref{eq:A1}) must be consistent with the following constraints
\begin{equation}
\partial_{t}\left(\hat{h}J(\hat{\vartheta},\hat{\varphi})
\cos\hat{\varphi}\right)=0,\quad
J(\hat{\vartheta},\hat{\varphi})=\hat{\vartheta}_{s}\hat{\varphi}_{n}-
\hat{\vartheta}_{n}\hat{\varphi}_{s},\quad\hat{b}_{t}=0,\label{eq:A3}
\end{equation}
where $J(\hat{\vartheta},\hat{\varphi})$ denotes the Jacobian for the transformation (\ref{eq:A2}), and the hat symbol indicates that arguments of hatted functions are taken at $\vartheta=\hat{\vartheta}$, $\varphi=\hat{\varphi}$.

Due to the arbitrariness of choosing the labeling coordinates $s$ and $n$, it is convenient to assign them so that equal areas on the sphere contain equal volumes:
\begin{equation}
Hdsdn=h\cos{\varphi}d\varphi d\vartheta,\label{eq:A4}
\end{equation}
where $H$ is some constant. Then, after integrating Eqs.~(\ref{eq:A3}) over time, we obtain that
\begin{equation}
\hat{h}=\frac{H}{J(\hat{\vartheta},\sin\hat{\varphi})},\quad\hat{b}=B(s,n).\label{eq:A5}
\end{equation}
Here, as before, in Sec.~\ref{sec2}, the constant $H$ means the average layer depth, while the time-independent function $B$ defines the initial buoyancy distribution.

Now, it is possible to apply the transformation (\ref{eq:A2}) directly to the Lagrangian (\ref{eq:A1}) and make it free from constraints. Using the relations (\ref{eq:A5}), and $d\vartheta/dt=\hat{\vartheta}_{t}$, as well as given that the Lagrangian is defined only with accuracy to total time derivative of an arbitrary function, we find
\begin{equation}
\mathcal{L}=-\frac{Ha^{2}}{2}\int\left(2a^{2}\Omega\hat{\vartheta}_{t}\sin^{2}\hat{\varphi}+
\frac{HB}{J(\hat{\vartheta},\sin\hat{\varphi})}\right)dsdn.\label{eq:A6}
\end{equation}

It is easy to see that if we introduce notations
\begin{equation}
\hat{x}=a\hat{\vartheta},\quad\hat{y}=a\sin\hat{\varphi},\label{eq:A7}
\end{equation}
and renormalize the labeling coordinates $s\rightarrow as$, $n\rightarrow an$, then the Lagrangian (\ref{eq:A6}) takes the form
\begin{equation}
\mathcal{L}=-\frac{H}{2}\int\left(2\frac{\Omega}{a}\hat{x}_{t}\hat{y}^{2}+
\frac{HB}{J(\hat{x},\hat{y})}\right)dsdn,\label{eq:A8}
\end{equation}
and coincides with the Lagrangian (\ref{eq:13}) in the inertial-free approximation.

This fact means that the requirement $|y/a|\ll1$, traditional for equatorial models when choosing the Coriolis parameterization (\ref{eq:4}), is redundant, in our case, and can be replaced by a much softer condition $|y/a|\leq1$.

\section{Turbulence of equatorial straight-line jets}\label{sec7}

Let us consider turbulence occurring in the narrow equatorial belt as a collective effect to be created by a statistical ensemble of thin temperature jets. To exclude the influence of all other factors except the beta effect, we restrict ourselves to the simplest model that ignores the bending of the jets and treats them as straight lines parallel to the equator.

The random field of velocity produced by such jets can be found from the condition of equatorial geostrophic balance, putting, for simplicity, the layer depth $h$ is constant. Then, being written in a dimensionless form, this condition looks like the relation
\begin{equation}
\varepsilon yu=-\frac{1}{2}b_{y},\label{eq:B1}
\end{equation}
between the zonal velocity component $u$ and the derivative of buoyancy $b$ over the meridional coordinate $y$. The small parameter $\varepsilon=L/a$ results from scaling to be expressed through the characteristic scale $L$ and the Earth's radius $a$.

Since the Fourier transform is traditional for studying spectra, it is reasonable to use it for reformulating (\ref{eq:B1}) as the boundary value problem
\begin{equation}
u_{k}=-\frac{1}{4\pi\varepsilon}
\int^{\infty}_{-\infty}b_{y}e^{-iky}dy,\quad
u|_{k=0}=\int^{\infty}_{-\infty}u(y)dy=0.\label{eq:B2}
\end{equation}
Thus the Fourier amplitude of velocity $u$ treated as a function of the wave number $k$ is subject to the first-order linear differential equation and the boundary condition. It is worth noting that this condition follows directly from the ergodicity hypothesis (equivalence of spatial and statistical averaging) and the assumption of the turbulence in question being a zero-mean random process.

By considering the sum of step functions as the buoyancy distribution
\begin{equation}
b=\sum\gamma_{i}\theta\left(y-y_{i}\right),\label{eq:B3}
\end{equation}
and putting it into (\ref{eq:B2}), after some algebra, we get
\begin{equation}
u=\frac{1}{2\pi\varepsilon}\sum\frac{\gamma_{i}}{y_{i}}
\sin\left(\frac{ky_{i}}{2}\right)e^{-iky_{i}/2}.\label{eq:B4}
\end{equation}
This expression describes the Fourier amplitude of the velocity field produced by an equatorial ensemble of thin temperature jets. Free model parameters, which include buoyancy jumps $\gamma_{i}$ and their distances $y_{i}$ from the equator, are listed by the index $i$ and, below, play the role of independent random variables.

Now, to calculate the energy density spectrum $E(k)$, it is necessary to use the standard procedure
\begin{equation}
E=\frac{1}{2}\langle |u(k)|^{2}\rangle,\label{eq:B5}
\end{equation}
where angle brackets imply averaging.

In particular, when $y_{i}$ has a uniform distribution in the narrow belt $|y_{i}|\leq1$, the calculation gives the result
\begin{equation}
E=\frac{N\langle\gamma^{2}\rangle}{2\varepsilon^{2}}\int_{0}^{1}\frac{1}{y^{2}}
\sin^{2}\left(\frac{1}{2}ky\right)dy\propto\cos{k}-1+k\Si(k),\label{eq:B6}
\end{equation}
where $N$ is the mean number of jets per unit width of the equatorial belt, $\langle\gamma^{2}\rangle$ is the mean square magnitude of buoyancy jumps, and $\Si(k)$ is the integral sinus.

Thus, in the case of the equatorial beta effect, the energy density spectrum has two parts with different power laws. At small wave numbers, if $k\ll1$, the spectral density behaves as $E\propto k^{2}$, and at large ones, if $k\gg1$, as $E\propto k$.

In conclusion, we note that turning off the equatorial beta effect, in other conditions being equal, leads to a rearrangement of the spectrum (\ref{eq:B6}), so after that, it converts in
\begin{equation}
E\propto1-\frac{\sin k}{k}.\label{eq:B9}
\end{equation}
Because of this, as shown in Fig.~\ref{fig4}, the short-wave part of the spectrum loses its slope.

\end{document}